\documentclass[%
  onecolumn
]{mpi2015-cscpreprint}
\usepackage[T1]{fontenc}
\usepackage[utf8]{inputenc}
\usepackage{subfigure}
\usepackage{amssymb}
\usepackage{amsthm}
\usepackage{amsmath}
\usepackage{cleveref}
\usepackage{graphicx}
\begin{document}

\title{Adjacency-based, Non-intrusive Reduced-order Modeling for
Fluid-Structure Interactions}

\author[$1$,$\ast$]{Leonidas Gkimisis}
\author[$2$]{Thomas Richter}
\author[$1,2$]{Peter Benner}

\affil[$1$]{Max Planck Institute for Dynamics of Complex Techincal Systems, Sandtorstra{\ss}e 1, 39106 Magdeburg, Germany,
Computational Methods in Systems and Control Theory (CSC).}

\affil[$2$]{Otto-von-Guericke Otto-von-Guericke-
Universit\"at Magdeburg, Universit\"atsplatz 2, 39106 Magdeburg, Germany, Faculty of Mathematics, Institute for Analysis and Numerics.}

\affil[$\ast$]{corresponding author, \email{gkimisis@mpi-magdeburg.mpg.de}}



  
\shorttitle{Adjacency-based, non-intrusive model reduction for VIV}
\shortauthor{L. Gkimisis, T. Richter, P. Benner}
\shortdate{}
  


\novelty{Non-intrusive reduced-order modeling method for two-way coupled FSI. Sparse, nonlinear data-driven model inference. Coupled FSI model for Hron-Turek FSI3 benchmark testcase.}

\abstract{%
Non-intrusive model reduction is a promising solution to system dynamics prediction, especially in cases where data are collected from experimental campaigns or proprietary software simulations. In this work, we present a method for non-intrusive model reduction applied to Fluid-Structure Interaction (FSI) problems. The approach is based on the a priori known sparsity of the full-order system operators, which is dictated by grid adjacency information. In order to enforce this type of sparsity, we solve a ``local'', regularized least-squares problem for each degree of freedom on a grid, considering only the training data from adjacent degrees of freedom, thus making computation and storage of the inferred full-order operators feasible. After constructing the non-intrusive, sparse full-order model, Proper Orthogonal Decomposition (POD) is used for its projection to a reduced dimension subspace and thus the construction of a reduced-order model (ROM). The methodology is applied to the challenging Hron-Turek benchmark FSI3, for Re = 200. A physics-informed, non-intrusive ROM is constructed to predict the two-way coupled dynamics of a solid with a deformable, slender tail, subject to an incompressible, laminar flow. Results considering the accuracy and predictive capabilities of the inferred reduced models are discussed.
}
\maketitle                   
\section{Introduction}%
\label{sec:intro}
Fluid-Structure Interactions encompass coupled, strongly nonlinear fluid-solid dynamics systems, which arise in various engineering fields. For different regimes, materials and geometry configurations, the specific coupling conditions and corresponding FSI dynamical response can widely differ \cite{Rich}. In this work, we focus on incompressible fluid dynamics at low Reynolds numbers, coupled with deformable solids, at relatively low mass ratios. Under such conditions, the timescales and inertia of the two subsystems are comparable, leading to a strong, two-way coupling mechanism. 

The development of non-intrusive, reduced-order models for FSI problems is motivated by needs in engineering design and control. Firstly, the high computational cost of direct FSI numerical simulation renders multi-query tasks such as design optimization and real-time control cumbersome. Such engineering tasks could benefit from model reduction techniques, which have proven to allow for considerable speed-ups of FSI simulation, without significant loss of accuracy \cite{RozzaFSI}. Secondly, in cases where experimental or FSI simulation data are available, without access to a numerical simulation source code, non-intrusive modeling methods are developed for system dynamics prediction \cite{niromFSI}. Exploiting knowledge on the structure of the underlying dynamical system has been shown to enhance the predictive capabilities of non-intrusive ROMs for specific FSI regimes such as aeroelasticity \cite{OpInfAero}.

In this work, we implement an adjacency-based approach to FSI problems. The method was recently studied for linear advection and diffusion problems in \cite{SchuInf} and \cite{pidmd} and has been applied to the inference of quadratic-bilinear, incompressible fluid dynamics operators in Vortex-Induced Vibrations problems by the authors \cite{VIVinf}. Two coupled, data-driven models are inferred for the fluid dynamics and solid dynamics subsystems, exploiting a physics-based structure of the governing equations under the Arbitrary Lagrangian-Eulerian (ALE) formulation. We present results for the FSI3 Hron-Turek benchmark \cite{bench}, which has been previously studied for intrusive model reduction in \cite{RozzaFSI}. Through this testcase, we examine the predictive capabilities of the obtained, non-intrusive model.

\section{Theoretical Background}
\label{sec:theory}
\subsection{Governing equations}
\label{subs:eqns}
For the fluid flow, we focus on the regime of 2D incompressible, laminar flows. In cases of moving fluid domains $\Omega(t)$, the equations can be expressed with respect to a reference configuration $\hat{\Omega}$ using the ALE formulation. The map from the reference domain to the current configuration is given as: $\hat{\mathbf{x}} \mapsto \mathbf{x}=\hat{\mathbf{x}}+\hat{\mathbf{d}}_f(t)$, where $\hat{\mathbf{d}}_f(t)$ is a deformation field. The incompressible ALE Navier-Stokes equations are then given by
\begin{equation} 
\label{NS_ALE}
\left\{\begin{array}{ll}
\begin{aligned}
\rho_{f} J\left(\partial_{t} \hat{\mathbf{u}}+\hat{\nabla} \hat{\mathbf{u}} \mathbf{F}^{-1}\left(\hat{\mathbf{u}}-\partial_{t} \hat{\mathbf{d}}_f\right)\right)
-\operatorname{div}\left(J \hat{\boldsymbol{\sigma}}\left(\hat{\mathbf{u}}, \hat{p}\right) \mathbf{F}^{-T}\right) & =J \rho_f \vec g \quad\\ 
\operatorname{div}\left(J \mathbf{F}^{-1} \hat{\mathbf{u}}\right) & =0 \quad
\end{aligned}
\end{array}\right.,
\end{equation}

\noindent where $\hat{\mathbf{u}}$ is the ALE velocity field, $\hat{p}$ is the pressure, $\hat{\boldsymbol{\sigma}}$ is the ALE stress tensor given by
\begin{equation}
\label{sigF}
 \hat{\boldsymbol{\sigma}}\left(\hat{\mathbf{u}}, \hat{p}\right) =-\hat{p}I+\rho_f \nu_f \left( \hat{\nabla} \hat{\mathbf{u}} \mathbf{F}^{-1}+\mathbf{F}^{-T} \hat{\nabla} \hat{\mathbf{u}}^T \right),
\end{equation}

\noindent $\mathbf{F}$ is the gradient of the ALE map and $J=\det{\mathbf{F}}$, with 
\begin{equation}
\label{jacF}
\mathbf{F}=I+\hat{\nabla} \hat{\mathbf{d}}_f.
\end{equation}

\noindent For solid dynamics modelling, we use the linear Navier-Lam\'e equations, assuming no material or geometrical nonlinearities. The boundary of the solid is split into $\Gamma_1^s$ where the displacement is zero and $\Gamma_2^s$ where the fluid-structure coupling occurs. The solid deformation $\mathbf{d}_s$ is thus modeled by
\begin{equation} 
\label{solid}
\left\{\begin{array}{ll}
\partial_{tt} \mathbf{d}_s-\nabla \cdot \boldsymbol{\sigma}_s=0\\
\boldsymbol{\sigma}_s=\mu(\nabla \mathbf{d}_s + \nabla \mathbf{d}_s^T) + \lambda \; tr(\frac{1}{2}(\nabla \mathbf{d}_s + \nabla \mathbf{d}_s^T))I
\end{array}\right..
\end{equation}

\begin{figure}[!htbp]
\centering
  \centering
    \includegraphics[width=0.45\textwidth]{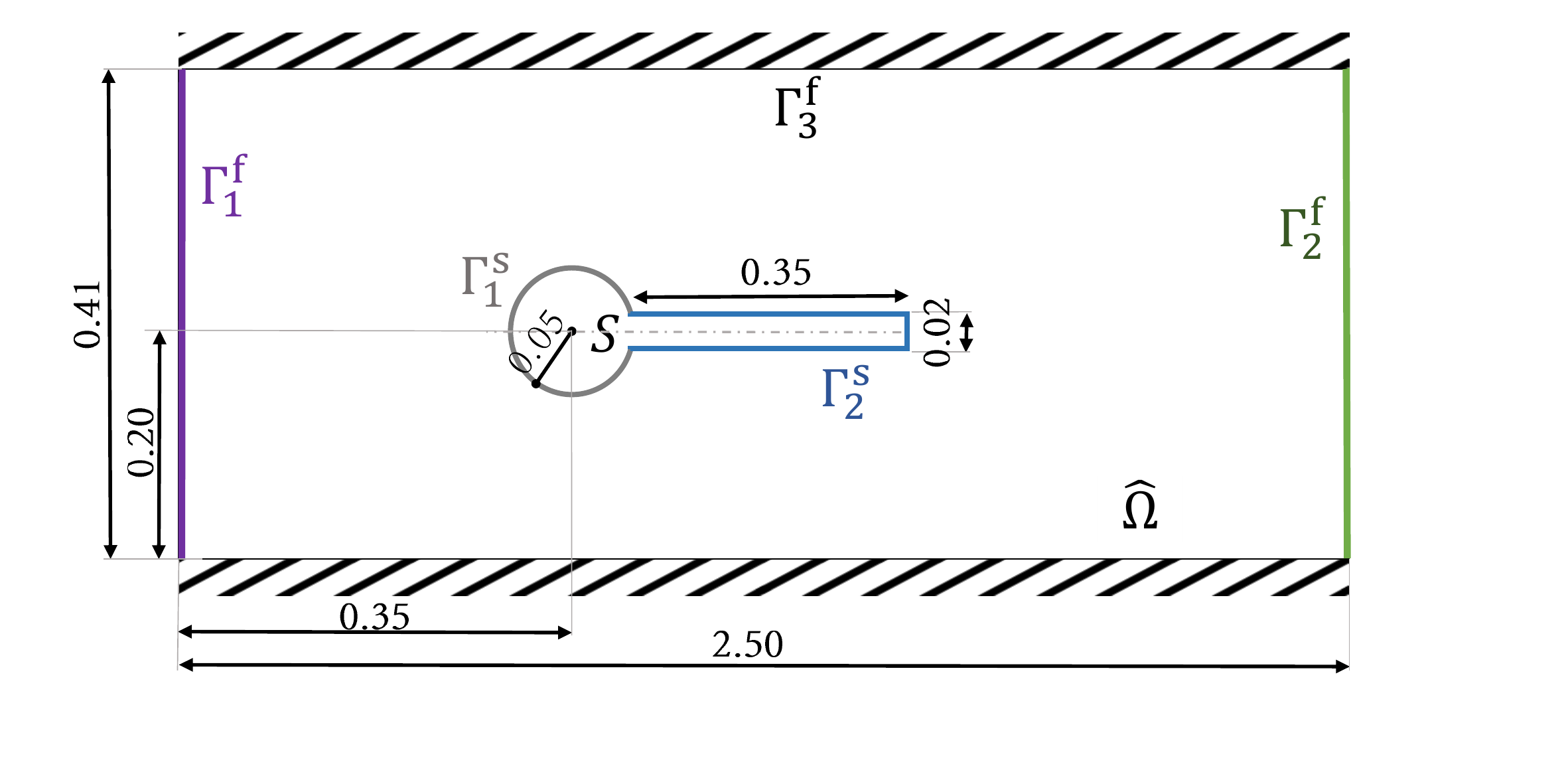}
  \captionof{figure}{Schematic representation (not scaled) of the Hron-Turek FSI benchmark testcase. A velocity profile is assigned at the inlet $\Gamma_1^f$ and FSI coupling occurs on the surface $\Gamma_2^s$ of a slender, deformable tail.}
  \label{fig:schematic}
\end{figure}

\noindent The boundary conditions of the coupled fluid-solid problem are given below, based on \Cref{fig:schematic}.
\begin{equation}
\label{BCs}
\left\{\begin{array}{ll}
\begin{aligned}
\hat{\mathbf{u}}&=\mathbf{u}_{in} \; & on \; \Gamma_1^f,\\
 J \left( \nu_f \hat{\nabla} \hat{\mathbf{u}} \mathbf{F}^{-1} -\hat{p}I  \right) \mathbf{F}^{-T} \cdot \hat{\mathbf{n}}_f &=\mathbf{0} \; & on \; \Gamma_2^f,\\
\hat{\mathbf{u}}&=\mathbf{0} \; & on \; \Gamma_3^f,\\
\mathbf{d}_s&=\mathbf{0} \; & on \; \Gamma_1^s,\\
J \hat{\sigma}_f \mathbf{F}^{-T} \cdot \hat{\mathbf{n}}_f &= \boldsymbol{\sigma}_s \cdot \mathbf{n}_s  \; & on \; \hat{\Omega} \cap \cal S,\\
\hat{\mathbf{u}}&=\partial_{t} \mathbf{d}_s \; & on \; \hat{\Omega} \cap \cal S.\\
\end{aligned}
\end{array}\right.
\end{equation}

\noindent The above formulation considers the coupled problem on a reference fluid domain $\hat{\Omega}$. We use a stiffened harmonic extension as an ALE map, to transfer the solution to the current, deformed configuration $\Omega(t)$.
\begin{equation} 
\label{extend}
- \hat{\nabla}^2 \hat{\mathbf{d}}_f - \frac{\hat{\nabla} a(\hat{\mathbf{x}})}{a(\hat{\mathbf{x}})} \hat{\nabla} \hat{\mathbf{d}}_f =\mathbf{0}, \; \text{in} \; \hat{\Omega},
\end{equation}

\noindent with boundary conditions on the interface $\hat{\Omega} \cap \cal S$, and on the domain external boundaries:
\begin{equation}
\label{lapBC}
\left\{\begin{array}{l}
\begin{aligned}
&\hat{\mathbf{d}}_f =\hat{\mathbf{d}}_s-\hat{\mathbf{d}}_s(t=t_1),  & \; \text{on} \; \hat{\Omega} \cap \cal S,\\
&\hat{\mathbf{d}}_f =\mathbf{0}, & \; \text{on} \; \Gamma_1^f \cup \Gamma_2^f \cup \Gamma_3^f.
\end{aligned}
\end{array}\right.
\end{equation}

\noindent $a(\hat{\mathbf{x}})$ introduces a stiffening effect close to the FSI interface $\hat{\Omega} \cap \cal S$, which allows for larger mesh displacements without element degradation, compared to a purely harmonic extension (see Section 5.3.5 of \cite{Rich}). $t_1$ is selected as a reference time, for which $\hat{\Omega}=\Omega(t_1)$.

\subsection{Approximate model structure}

By discretizing \cref{NS_ALE}, \cref{solid} and \cref{extend}, we aim to reveal an approximate structure of the dynamical system, which will be exploited for non-intrusive modeling. We assume that the gradient of the deformation field $\hat{\nabla} \hat{\mathbf{d}}_f$ is sufficiently small, such that $F \approx I$ (and thus $J \approx 1$) is a valid approximation of \cref{sigF}. In that case, the structure of \cref{NS_ALE} is approximately quadratic-bilinear. After discretizing in space and time, working similarly to \cite{BennNS}, we cancel out the algebraic contribution of the pressure and come up with the following, approximate model structure for the fluid velocity field:
\begin{equation} 
\label{quadbil}
\mathbf{u}^{k+1} \approx A \mathbf{u}^k + H (\mathbf{u}^{k+1})^2 + K \; \partial_{t} \mathbf{d}_f \otimes \mathbf{u}^{k+1} 
 + B \; \partial_{t} \mathbf{d}_s + L \; \mathbf{u}_{in} + C,
\end{equation}

\noindent where $(\mathbf{u})^2$ denotes the vector of $2n_f \choose 2$ unique elements of the Kronecker product  $\mathbf{u} \otimes \mathbf{u}$.
For the solid velocity field, we get a second-order, forced linear system by discretizing \cref{solid} in space and time
\begin{equation} 
\label{sollin}
\partial_{t} \mathbf{d}_s^{k+1}=A_s \partial_{t} \mathbf{d}_s^{k} + K_s \; \mathbf{d}_s^{k}+\mathbf{f}.
\end{equation}

\noindent The forcing $\mathbf{f}$ originates from the dynamic coupling condition on the interface (i.e. the projection of the fluid stress tensor on the normal to the interface). This can be approximated by a quadratic model, as follows.
\begin{equation} 
\label{forcing}
\mathbf{f} \approx B_s \mathbf{u}_{FS}^{k+1}+H_s {\mathbf{u}_{FS}^{k+1}}^2,
\end{equation}

\noindent where $FS$ denotes nodes that are sufficiently close to the interface, given the sparsity of the discretized stress tensor (see \cref{sigF}). This means that $\mathbf{u}_{FS}=R \mathbf{u}$, where $R$ is an ordered selection matrix.

\noindent Finally, the solid velocity field $\partial_{t} \mathbf{d}_s$ should be integrated to update the solid displacement field $ \mathbf{d}_s$. To this end, a Crank-Nicholson scheme is used
\begin{equation}
\label{crnic}
    \mathbf{d}_s^{k+1}=\mathbf{d}_s^{k}+\frac{\Delta t}{2}(\partial_t \mathbf{d}_s^{k+1}+\partial_t \mathbf{d}_s^{k}).
\end{equation}

\noindent We also need to discretize and solve \cref{extend}, in order to compute the fluid mesh displacement in the current configuration, $\Omega(t)$, with respect to the reference configuration $\hat{\Omega}$. This leads to 
\begin{equation}
\label{ext_disc}
\mathbf{d}_f^{k}=A_L S \mathbf{d}_{s}^{k},
\end{equation}

\noindent where $A_L$ is the inverse of the discretized ALE map and $S$ is an ordered selection matrix, which picks only the solid DoFs on $\Gamma_2^s$.
\noindent Equations \eqref{quadbil} to \eqref{ext_disc} comprise a discretized, coupled dynamical system, which models ALE FSI problems under small deformations. In the following, we use this structure to fit the corresponding operators to numerical FSI data and thus construct a physics-based, non-intrusive model.

\section{FSI Non-Intrusive Model Reduction}
\label{sec:ROM}
\subsection{Physics-based adjacency}
\label{subs:pba}

Having identified an approximate structure for the FSI dynamical system, we aim to infer the unknown operators of the fluid dynamics subsystem (\cref{quadbil}), the solid dynamics subsystem (\cref{sollin}) and the solid forcing term (\cref{forcing}), from simulation data. 

The discretized operators in \cref{sollin} are sparse, while the operators of \cref{quadbil} and \cref{forcing} are only approximately sparse, due to canceling out the contribution from the pressure. The corresponding sparsity pattern originates from grid adjacency. Based on this observation, the employed method aims to infer full-order, adjacency-based, sparse operators.

To proceed with the inference task, we should map the available data for the flowfield over time $t=[0,T]$ to a reference configuration $\hat{\Omega}$, since equation \cref{quadbil} was derived there. Knowing the motion of the FSI interface from the data, we define a reference time as $t_1=\text{argmin}(\| \hat{\mathbf{d}}_{s}(t) \|_2 )$. Based on the reference position of the FSI interface at $t_1$, we can construct a fluid and a solid mesh (e.g. via Delaunay triangulation), with the constraint that the nodes of the two meshes on the interface should match. 

For the fluid, we compute the deformed mesh $\Omega(t)$ from \cref{ext_disc}, given the motion of the FSI interface (see \cref{lapBC}) and the constructed reference $\hat{\Omega}=\Omega(t_1)$. Then, the fluid velocity data can be interpolated on the reference configuration $\hat{\Omega}$ at each of the $n_T$ timesteps. For the solid, we interpolate the displacement field values for all $n_T$ timesteps from the solver mesh configuration at $t_1$, to the constructed solid mesh. The difference between handling the solid and the fluid problem should be noted: Mesh construction is necessary only for the fluid side due to the ALE formulation. However, grid adjacency information for both the fluid and the solid mesh is essential for the following. Thus, the solid mesh of the simulation data can be readily used as long as adjacency information can be retrieved.

\subsection{Least Squares formulation and regularization}

The number of grid nodes is denoted by $n_f$ for the fluid and $n_s$ for the solid mesh. Then, there are $2n_f$ and $2n_s$ kinematic degrees of freedom (DoFs), respectively, for this 2D problem. For the $i$-th DoF (streamwise or transverse component of $\mathbf{u}$ or $\partial_{t}\mathbf{d}$), we denote by $q_A(i)$ the set of geometrically adjacent DoFs in the fluid $A\equiv f$ or solid $A\equiv s$ mesh respectively. For example, if $i$ is a fluid velocity component, $q_f(i)$ includes exclusively DoFs which correspond to fluid mesh nodes, adjacent to the node of $i$. Then, the local inference of the operators for DoF $i$ corresponds to solving the following problem:
\begin{equation} 
\label{LS_noreg}
\min _{\boldsymbol{\beta}_i}\left\| \boldsymbol{\beta}_i^T \mathcal{D}_i-\mathbf{y}_i^{m+1}\right\|_{2},
\end{equation}

\noindent where we make a distinction between the fluid and solid dynamics problems:

\noindent For the fluid dynamics subsystem,
\begin{equation}
\label{LSfl}
\boldsymbol{\beta}_i=[C_i,A_{i,.}, H_{i,.}, K_{i,.}]^T, \;\;
\mathcal{D}_i^T=
\begin{bmatrix}
  \mathbf{1}\\
  \mathbf{u}_{q_f(i)}^m \\
  {\mathbf{u}_{q_f(i)}^m}^2\\
  \mathbf{u}_{q_f(i)}^m \otimes \partial_t \hat{\mathbf{d}}_f^m\\
\end{bmatrix},
\;\;
\mathbf{y}_i^{m+1}=[u_i^{t=2} \; ... \; u_i^{t=n_{T_1}}]^T,
\end{equation}

\noindent where $T_1$ corresponds to the training time and $m=[1,...,n_{T_1}-1]$. For fluid DoFs on $\Gamma_1^f$ and $\hat{\Omega} \cap \cal{S}$
we can directly impose the corresponding Dirichlet boundary conditions, by setting row $i$ of $A,H,K,C$ to zero, and assigning entries of value $1$ in corresponding positions of $L$ and $B$, accordingly.

\noindent For the solid dynamics subsystem, we introduce sets $q_{FS}(i)$ and $q_{I}(i)$, to distinguish between internal DoFs and DoFs on $\Gamma_2^s$. As a result, $q_{FS}(i)=q_f(i)$ for DoFs $i$ on $\Gamma_2^s$ and $q_{FS}(i)=\emptyset$ otherwise, while $q_{I}(i)=q_s(i)$ everywhere, except for $\Gamma_2^s$, where $q_{I}(i)=\emptyset$. Then the variables of \cref{LS_noreg} for the local inference of the solid dynamics operators are
\begin{equation}
\label{LSsol}
\boldsymbol{\beta}_i=[A_{S_{i,.}},K_{S_{i,.}}, B_{S_{i,.}}, H_{S_{i,.}}]^T, \;\;
\mathcal{D}_i^T=
\begin{bmatrix}
  \partial_{t} \mathbf{d}_{s_{q_I(i)}}^k\\
  \mathbf{d}_{s_{q_I(i)}}^k\\
  \mathbf{u}_{q_{FS}(i)}^k\\
  {\mathbf{u}_{q_{FS}(i)}^k}^2\\
\end{bmatrix},
\;\;
\mathbf{y}_i^{m+1}=[\partial_{t} {d}_{s_i}^{t=2} \; ... \; \partial_{t} {d}_{s_i}^{t=n_{T_1}}]^T.
\end{equation}

\begin{figure}[!htbp]
\centering
\begin{minipage}{.53\textwidth}
  \centering
    \subfigure{\includegraphics[width=0.49\textwidth]{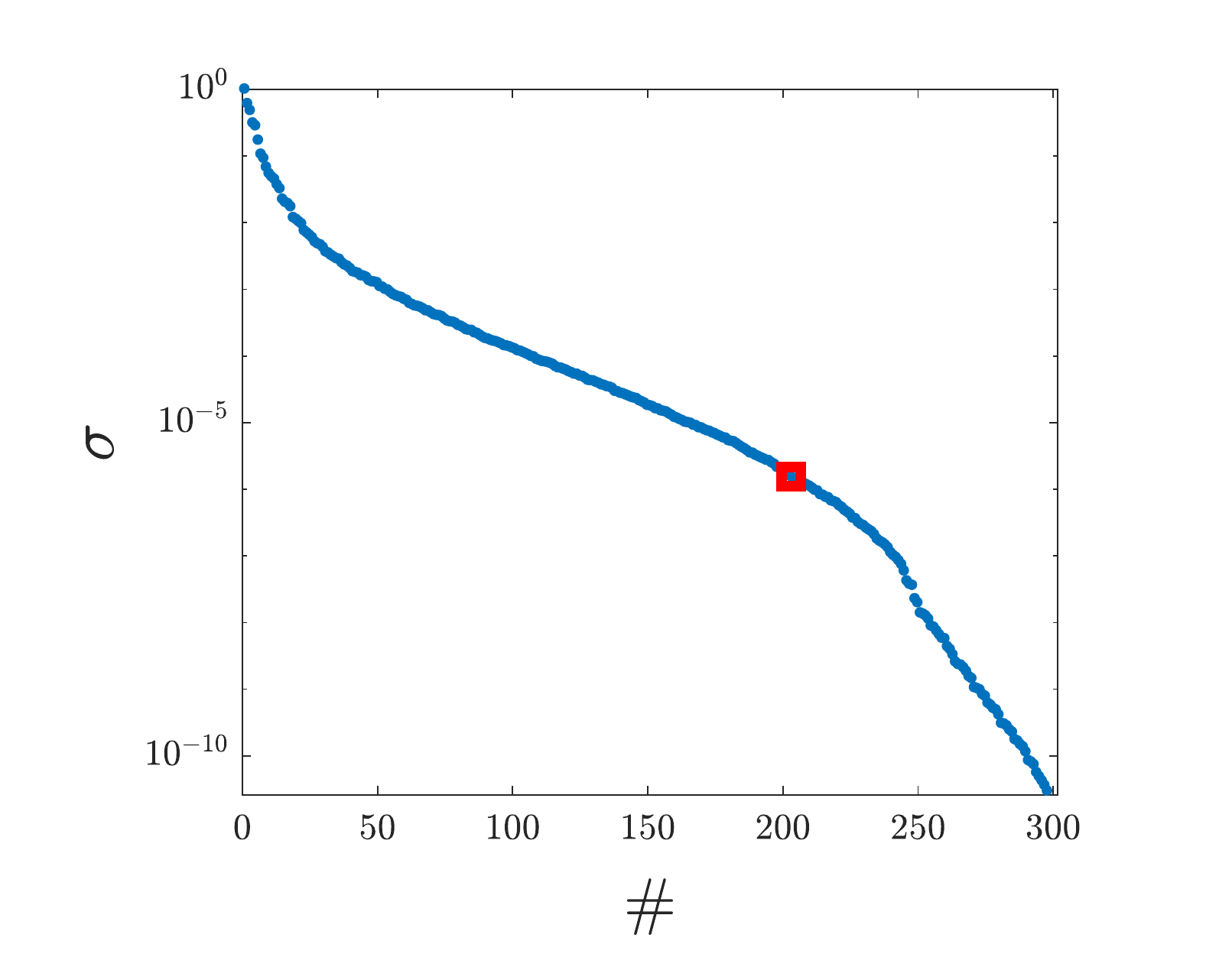}} 
    \subfigure{\includegraphics[width=0.49\textwidth]{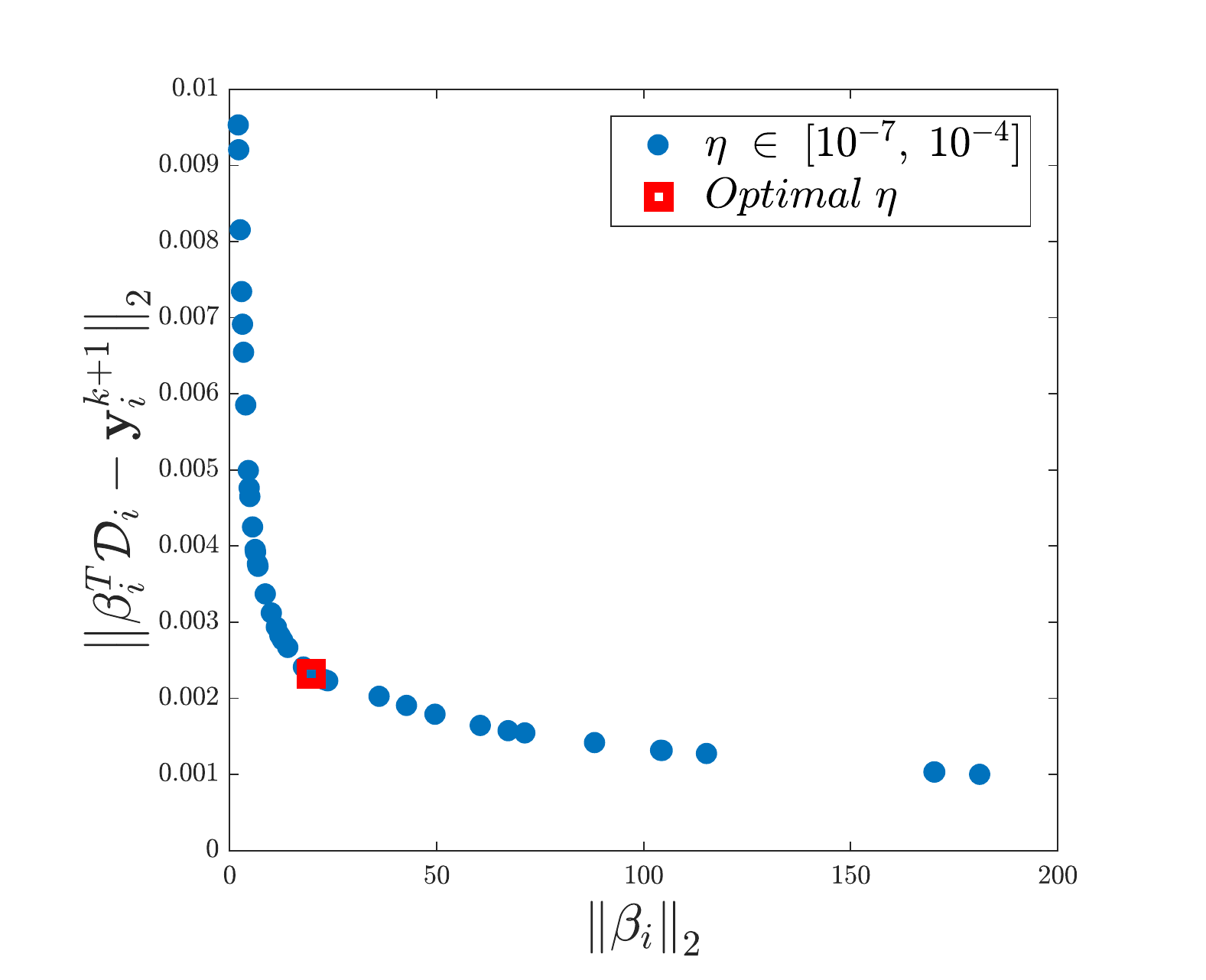}} 
  \captionof{figure}{Left: Truncation of the singular values of $\mathcal{D}$ to threshold $\eta$, for a random DoF $i$. Right: Corresponding L-curve: The optimal $\eta$ achieves a compromise between the least-squares error $\left\| \boldsymbol{\beta}_i^T \mathcal{D}_i-\mathbf{y}_i^{m+1}\right\|_{2}$ and the solution norm  $\left\| \boldsymbol{\beta}_i \right\|_{2}$.}
  \label{fig:eta}
\end{minipage}%
\hfill
\begin{minipage}{.43\textwidth}
  \centering
  \includegraphics[width=1.05\linewidth]{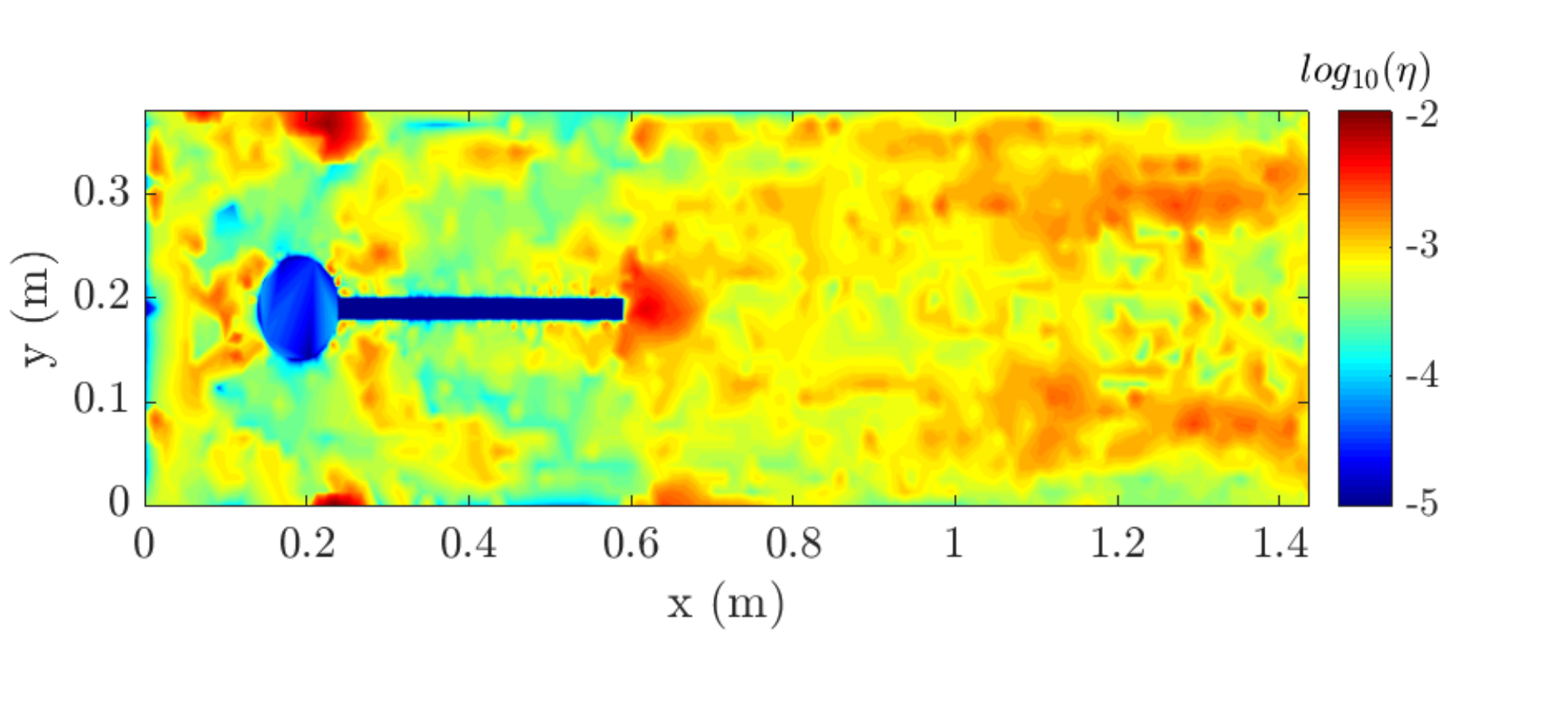}
  \captionof{figure}{Interpolated, sub-optimal $\eta$ value for the streamwise flowfield component, using the optimal $\eta$ for 30 \% of the DoFs. Regularization values are higher along the wake of the flow and close to the deformable solid tail.}
  \label{fig:test2}
\end{minipage}
\end{figure}

In practice, \cref{LS_noreg} should be complemented with a regularization term, to bypass potential numerical errors due to small singular values of $\mathcal{D}_i$. To this end, we employ truncated SVD of $\mathcal{D}_i$, for different truncation thresholds $\eta$ of the normalized singular values of $\mathcal{D}_i$. The optimal $\eta$ is determined as the ``corner'' of an L-curve, with the solution norm $\left\| \boldsymbol{\beta}_i \right\|_{2}$ on the $x$ axis and the error $\left\| \boldsymbol{\beta}_i^T \mathcal{D}_i-\mathbf{y}_i^{m+1}\right\|_{2}$ on the $y$ axis (see \Cref{fig:eta}). In essence, the SVD truncation corresponds to an $L_2$ regularization of \cref{LS_noreg}. Regularization plays a crucial role on the inferred model properties and thus determining the range of possible $\eta$ is not trivial. The process of computing the optimal threshold $\eta$ is performed for few, randomly sampled DoFs. An interpolation from the sampled, optimal $\eta$ values is performed to the rest of the mesh DoFs, as shown in \Cref{fig:test2}, such that only one least-squares problem per $i$ is solved for those, with a predefined truncation threshold.

\subsection{Model reduction and coupling}

After solving \cref{LS_noreg} for both the solid and the fluid nodes, we have inferred the sparse, full-order matrices of \cref{quadbil} and \cref{sollin}. In order to derive a reduced-order model, we use a POD projection. This is done by taking the SVD of the velocity flowfield $\mathbf{u}$ snapshot matrix, and of the solid velocity field $\partial_{t} \mathbf{d}_s$ snapshot matrix, over some training interval $[0,T_1]$. By retaining the first $r_f$ and $r_s$ singular modes accordingly, we obtain the orthonormal projection bases $\Phi_f \in \mathbb{R}^{2n_f \times r_f}$ and $\Phi_s \in \mathbb{R}^{2n_s \times r_s}$. Substituting the projected states $\mathbf{u}=\Phi_f \tilde{\mathbf{u}}$ and $\partial_{t} \mathbf{d}_s=\Phi_f \partial_{t} \tilde{\mathbf{d}}_s$ to \cref{quadbil} and \cref{sollin} results to a coupled, non-intrusive, reduced-order model

\begin{equation} 
\label{quadbil2}
\left\{\begin{array}{ll}
\begin{aligned}
\tilde{\mathbf{u}}^{k+1} &= \tilde{A} \tilde{\mathbf{u}}^k + \tilde{H} \tilde{\mathbf{u}}^{k+1} \otimes \tilde{\mathbf{u}}^{k+1} + \tilde{K} \; \partial_{t} \tilde{\mathbf{d}}_s \otimes \tilde{\mathbf{u}}^{k+1} 
 + \tilde{B} \; \partial_{t} \mathbf{d}_s + \tilde{L} \; \mathbf{u}_{in} + \tilde{C} \\
\partial_{t} \tilde{\mathbf{d}}_s^{k+1} &=\tilde{A}_s \partial_{t} \tilde{\mathbf{d}}_s^k + \tilde{K}_s \; \tilde{\mathbf{d}}_s^{k}+\tilde{B}_s \tilde{\mathbf{u}}^{k+1}+\tilde{H}_s \tilde{\mathbf{u}}^{k+1} \otimes \tilde{\mathbf{u}}^{k+1} \\
\tilde{\mathbf{d}}_s^{k+1} &=\tilde{\mathbf{d}}_s^{k}+\frac{\Delta t}{2}(\partial_t \tilde{\mathbf{d}}_s^{k+1}+\partial_t \tilde{\mathbf{d}}_s^{k}).
\end{aligned}
\end{array}\right.
\end{equation}

\noindent The corresponding ROM matrices of \cref{quadbil2} can be efficiently computed since the sparsity patterns of the full-order matrices are known. We then have for the fluid subsystem 
\begin{equation}
\label{projF}
\tilde{A}=\Phi_f^T A \Phi_f, \; \tilde{H}=\Phi_f^T H \Phi_f \otimes \Phi_f, \; \tilde{K}=\Phi_f^T K (A_L S \Phi_s) \otimes \Phi_f, \;
\tilde{B}=\Phi_f^T B \Phi_s, \; \tilde{L}=\Phi_f^T L, \; \tilde{C}=\Phi_f^T C,
    \end{equation}

\noindent and for the solid 
\begin{equation}
\label{projS}
\tilde{A}_s=\Phi_s^T A_s \Phi_s, \; \tilde{K}_s=\Phi_s^T K_s \Phi_s, \; 
\tilde{B}_s=\Phi_s^T B_s R \Phi_f, \; \tilde{H}_s=\Phi_s^T H_s (R \Phi_f) \otimes (R \Phi_f) \;.
\end{equation}

It was observed that the stiff coupling exhibited in FSI numerical simulation \cite{bench} transfers also to the data-driven, reduced-order model. The implicit formulation of \cref{quadbil} is necessary for stability. The combined convergence criterion for each timestep ${\left\| \tilde{\mathbf{u}}_{j+1}^k -\tilde{\mathbf{u}}_{j}^k\right\|}_2<res_1$ and ${\left\| \tilde{\mathbf{d_s}}_{j+1}^k -\tilde{\mathbf{d_s}}_{j}^k\right\|}_2<res_2$ was used during simulation, where $j$ denotes the index of iterations within timestep $k$. For each implicit iteration, a successive under-relaxation method was used.

\section{Numerical Results}

The analyzed methodology was tested for the Hron-Turek FSI3 benchmark \cite{bench}. Simulation data were obtained using the Gascoigne 3D open-source, Finite Element solver \cite{gas}. A monolithic approach was used, with a moderate mesh of $4128$ elements and a timestep of $\Delta t=5\times 10^{-3} \; s$, for a total time of $5 \; s$. We discard the first $1.5 \; s$ of data to avoid the influence of numerical solution initialization. The clock time required for the numerical simulation is approximately $1.5$ hours on a laptop.

As presented in \Cref{subs:pba}, we interpolate the data to a new mesh with $4060$ nodes, for which we construct the ALE map \cref{ext_disc}. We use $65\%$ of the overall available time series ($T_1=2.3 \;s$) to solve \cref{LS_noreg}, computing the optimal SVD truncation value for $10 \% $ of the mesh nodes, using $10$, logarithmically sorted values of $\eta \in [10^{-5},10^{-1}]$. We then construct the ROM through \cref{projF}, \cref{projS}, by selecting $r_f=40$, $r_s=5$, and solve the implicitly coupled, reduced-order system (\ref{quadbil2}) by successive under-relaxation and $res_1=res_2=10^{-6}$. The ``offline'' step of mesh construction, interpolation and FOM inference requires approximately $5$ minutes, while the simulation time of the ROM requires less than a minute.

\begin{figure}[!htb]
    \centering
    \subfigure[]{\includegraphics[width=0.43\textwidth]{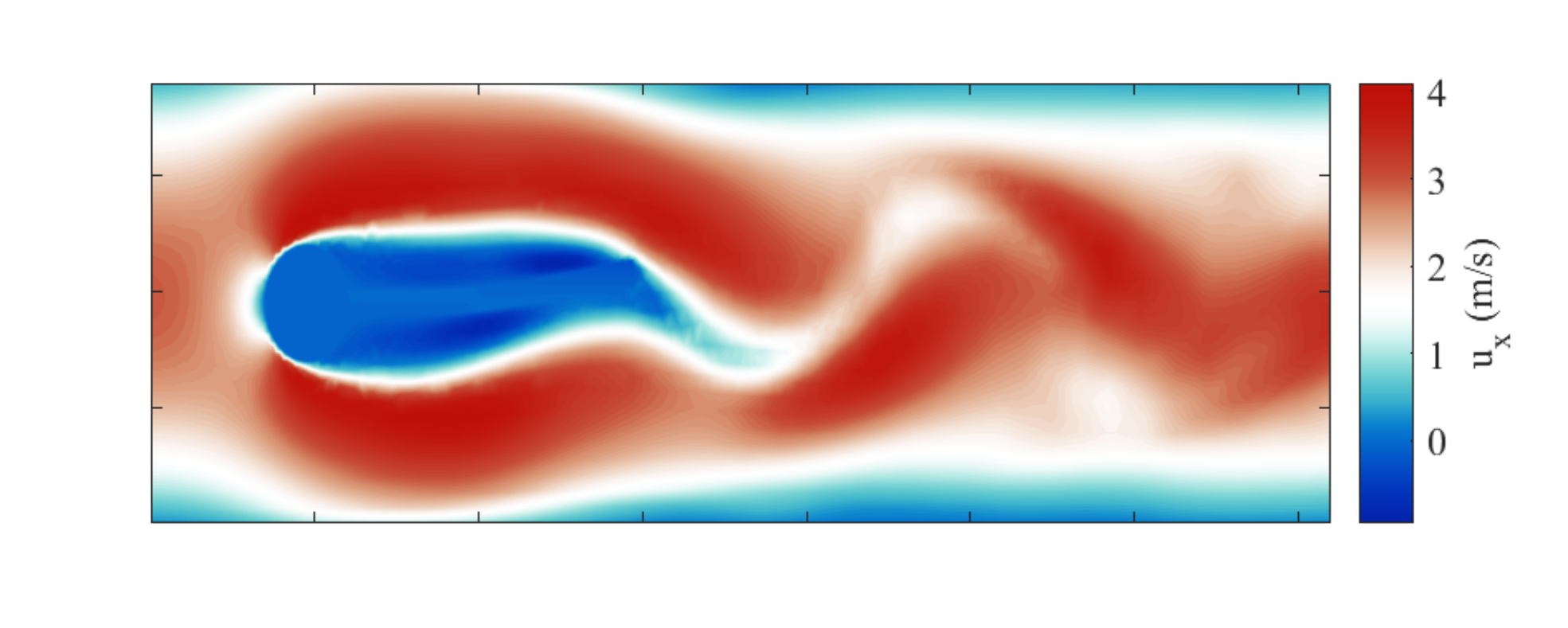}} 
    \subfigure[]{\includegraphics[width=0.43\textwidth]{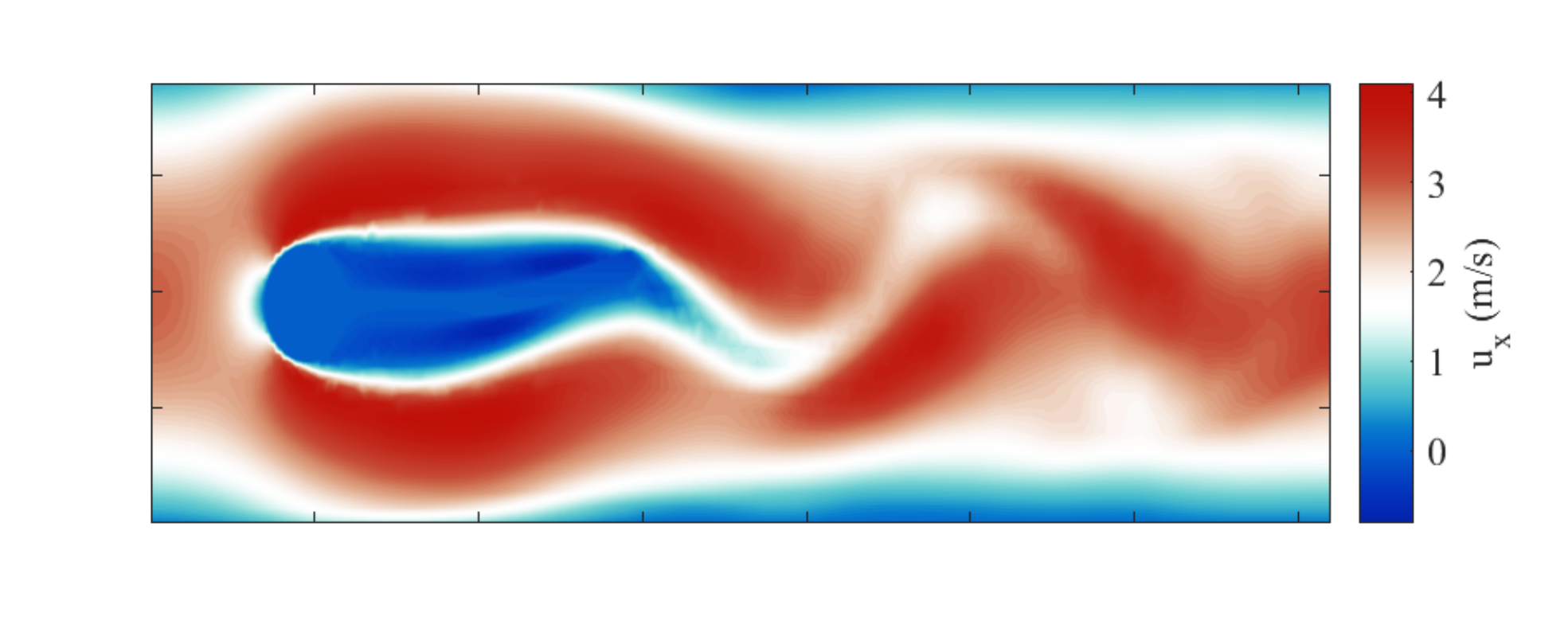}} 
    \subfigure[]{\includegraphics[width=0.43\textwidth]{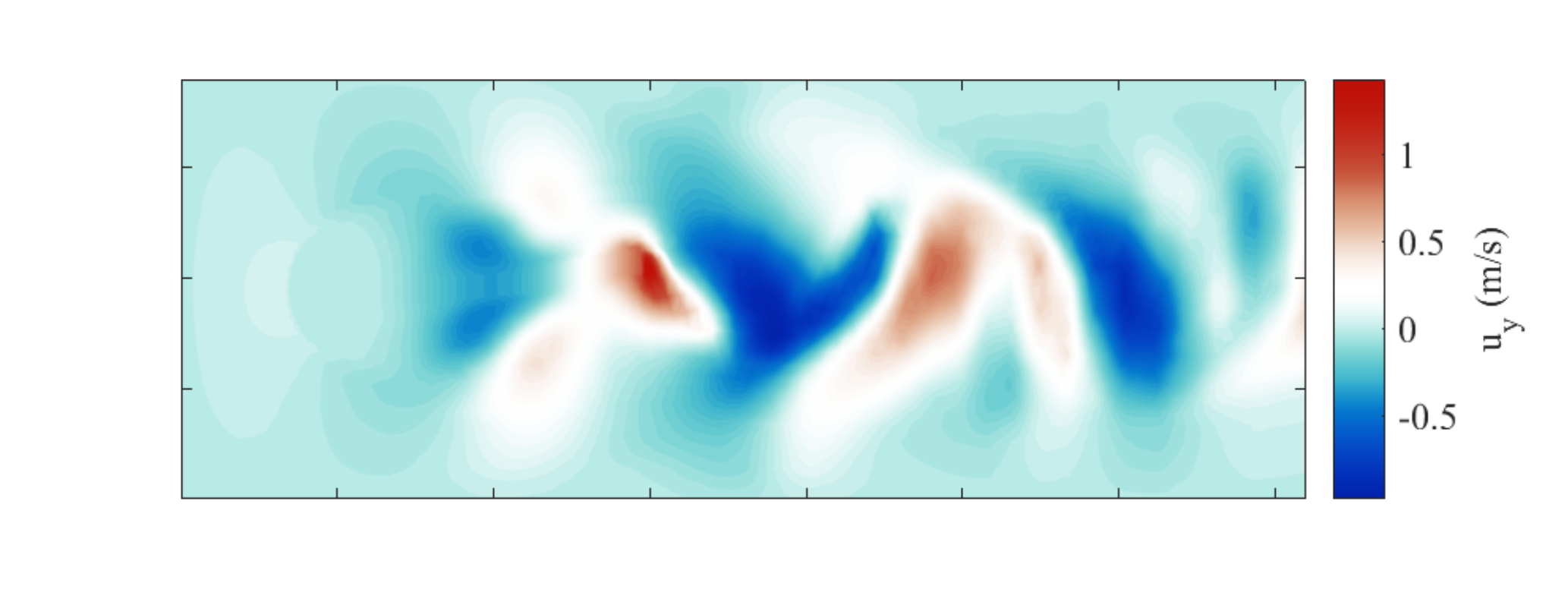}}
    \subfigure[]{\includegraphics[width=0.43\textwidth]{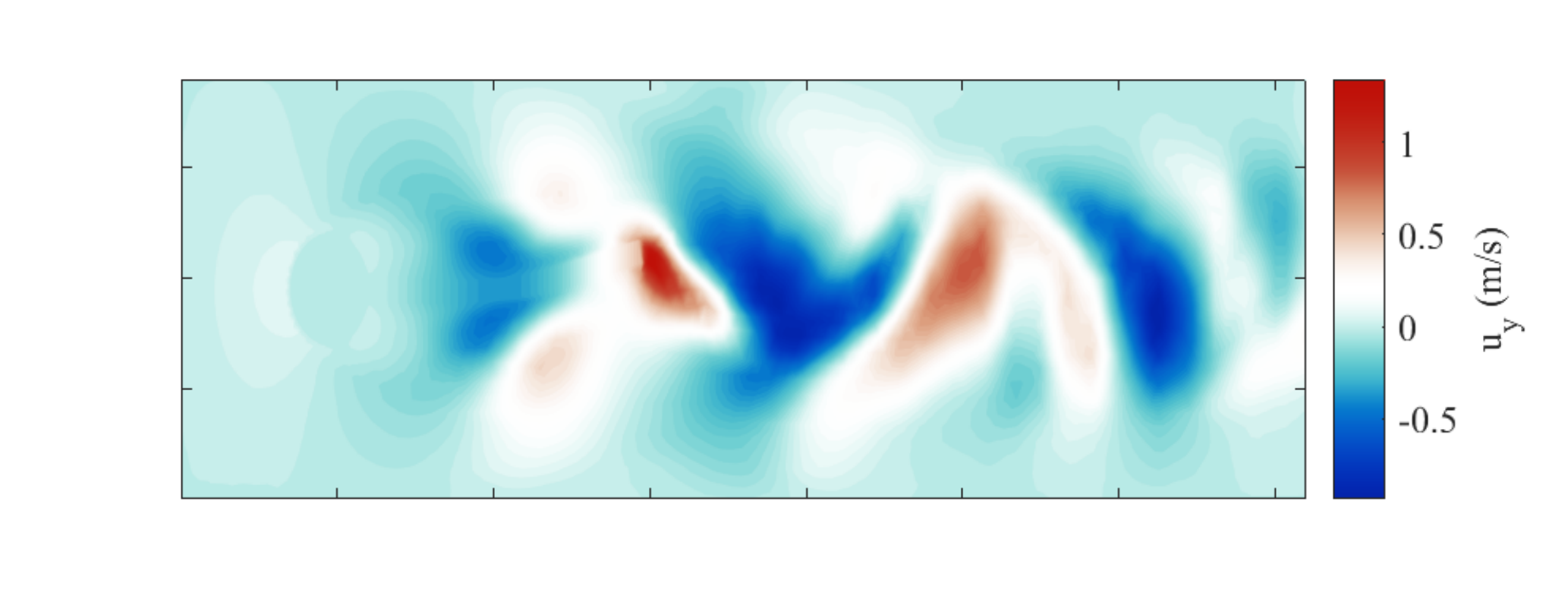}}
    \caption{(a) CFD results at $t=3.5\;s$ for streamwise velocity component (b) ROM results ($r_f=40$, $r_s=5$) at $t=3.5\;s$ for streamwise velocity component (c) CFD results at $t=3.5\;s$ for transverse velocity component (d) ROM results ($r_f=40$, $r_s=5$) at $t=3.5\;s$ for transverse velocity component}
    \label{fig1:ROMres}
\end{figure}

\begin{figure}[!htb]
\centering
\begin{minipage}{.47\textwidth}
  \centering
    \includegraphics[width=0.99\textwidth]{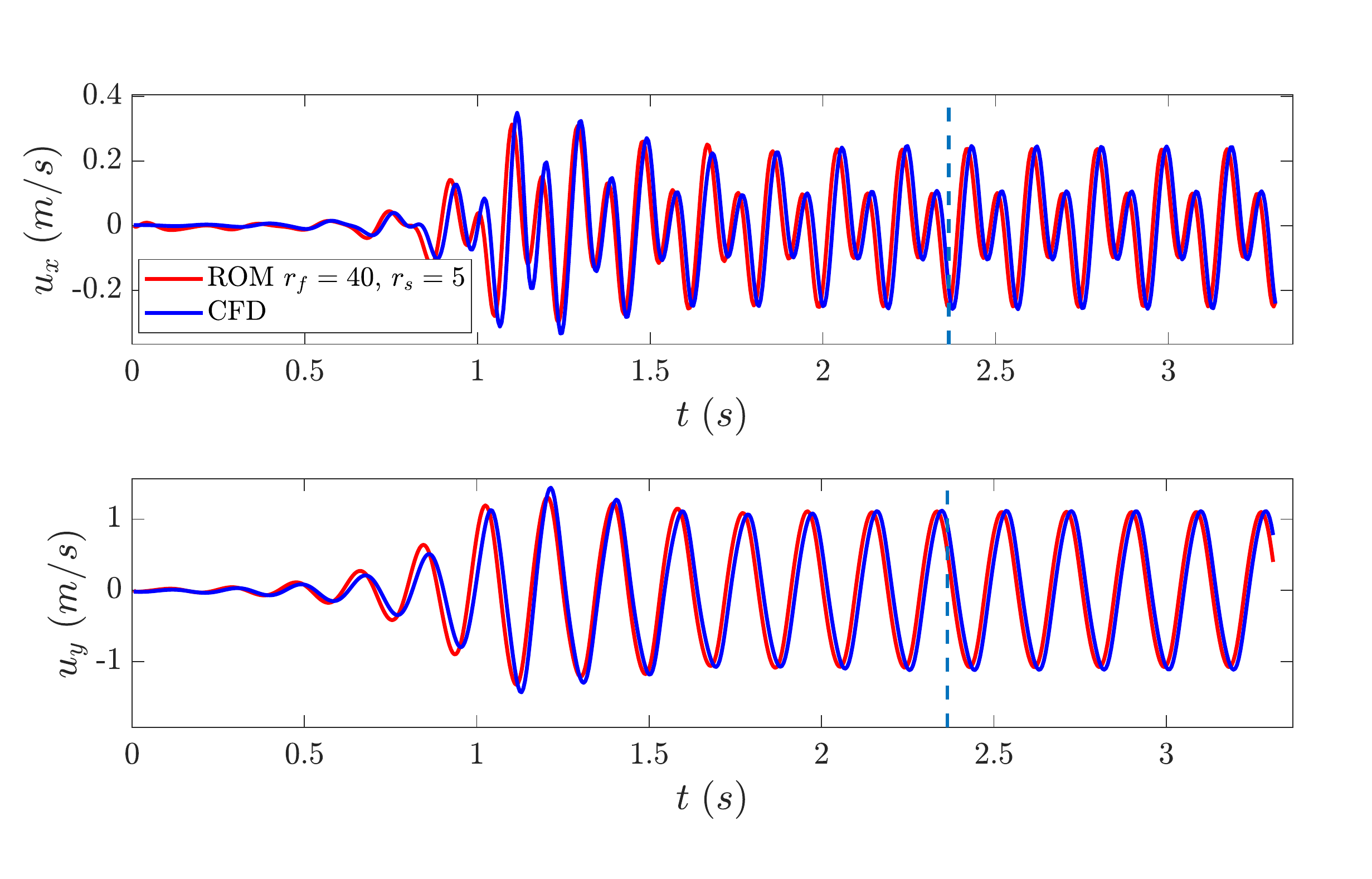}
  \captionof{figure}{Velocity ROM prediction (streamwise, transverse) for the upper right corner of the deformable solid. Training time $T_1=2.3\;s$ is denoted with a dashed vertical line.}
  \label{fig:osci}
\end{minipage}%
\hfill
\begin{minipage}{.51\textwidth}
  \centering
  \includegraphics[width=.70\linewidth]{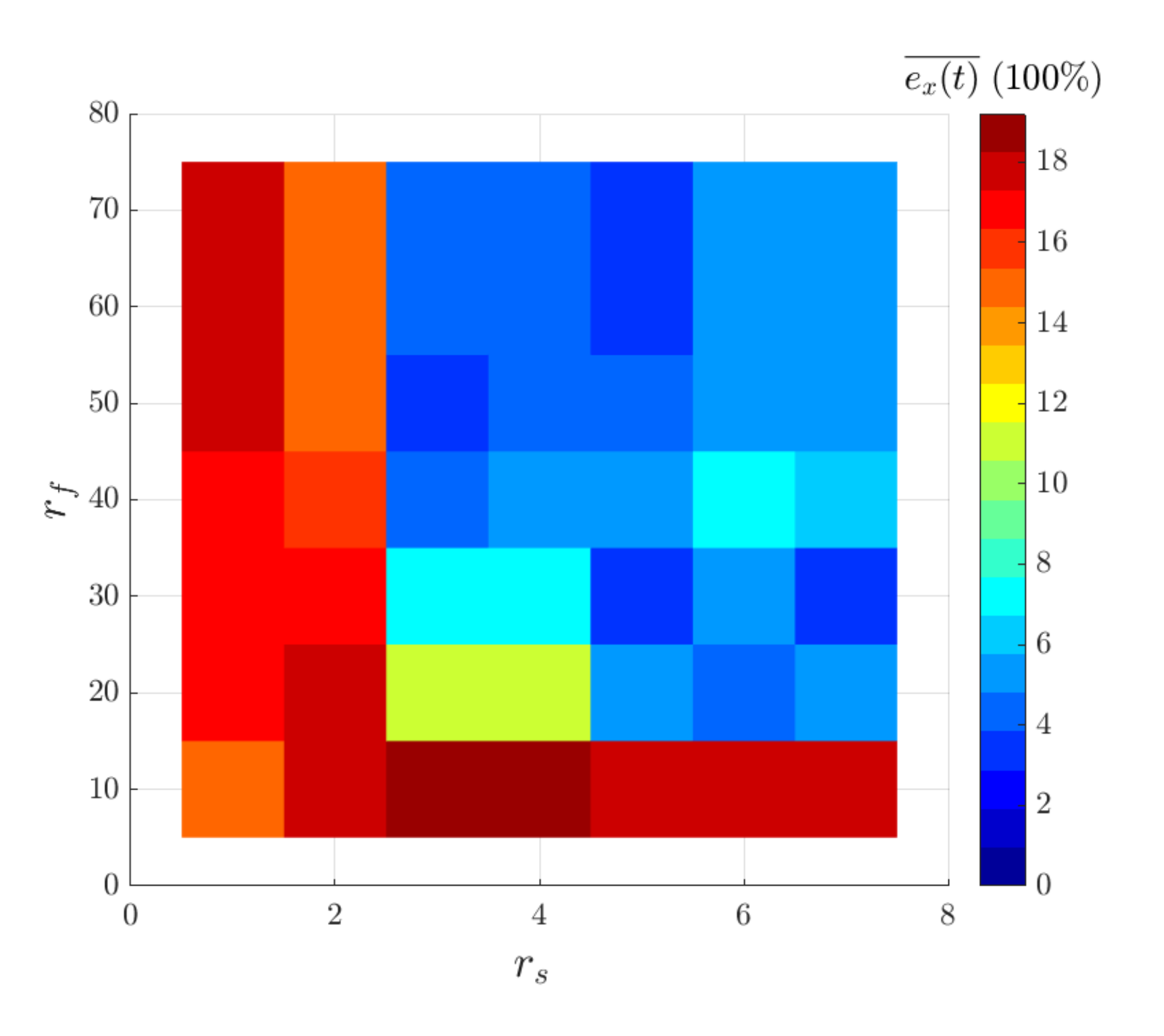}
  \captionof{figure}{Average error for streamwise flowfield over testing time $\overline{e_x(t)}$, for different combinations of ROM dimensions ($r_f,\;r_s$).}
  \label{fig:comp}
\end{minipage}
\end{figure}

The predicted, reconstructed ROM flowfield at $T=3.5\;s$, well beyond the training time $T_1$, is given in \Cref{fig1:ROMres}, in comparison to the CFD results for the same instance, for both velocity components. It is observed that the main dynamical features of vortex shedding as well as the motion of the deformable tail are well approximated, using only $45$ reduced variables, from the initial $8120$ DoFs. The vertical velocity over time for the upper right corner of the deformable tail is also presented in \Cref{fig:osci}. The predicted ROM solid motion is well in line with the numerical results and captures the limit cycle behaviour beyond training time $T_1$. A minor phase shift can be attributed to the considerable presence of transient dynamics in the training data.

Finally, it is interesting to investigate the stability of the derived FSI model with respect to the ROM dimensions $r_s$ and $r_f$ of the solid and fluid subsystems, respectively. In \Cref{fig:comp}, we illustrate the ROM streamwise velocity error with respect to the numerical simulation on the reference configuration $\hat{\Omega}$, averaged over the domain $\hat{\Omega}$ and testing time $t=[T_1,\;T]$. This is denoted with $\overline{e_x(t)}$. Based on $\overline{e_x(t)}$, we observe that $r_s\geq3$ modes are required for the solid subsystem for an accurate FSI ROM, as anticipated by the deformable tail motion. For values $r_s\geq 8$, a coupling instability was exhibited, irrespective of the $r_f$ value. Similarly $r_f\geq20$ SVD modes are necessary on the fluid side. However, we also observe that the combination of $r_f$ and $r_s$ can significantly affect the ROM predictive accuracy, due to the strong coupling on the FSI interface. For a significant range of ROM dimensions, the average error $\overline{e_x(t)}$ is below $5 \%$.

\section{Conclusions}
In this work, we employed an adjacency-based method for non-intrusive, reduced-order modelling of FSI problems. 
Accurate results were obtained for the Hron-Turek FSI3 benchmark testcase, with an average ROM prediction error of less than 5 \%. Based on the potential of this framework, aspects of computational efficiency, parametric model reduction, as well as as a theoretical analysis of local non-intrusive inference comprise potential, future research directions.

\vspace{\baselineskip}

\end{document}